# Ultrafast electron localization in a correlated metal

Jose R. L. Mardegan[1,2,*], Serhane Zerdane[3,*], Giulia Mancini[3,4], Vincent Esposito[3], Jeremy Rouxel[4], Roman Mankowsky[3], Cristian Svetina[3], Namrata Gurung[5,6], Sergii Parchenko[1], Michael Porer[1], Bulat Burganov[7], Yunpei Deng[3], Paul Beaud[3], Gerhard Ingold[3], Bill Pedrini[3], Christopher Arrell[3], Christian Erny[3], Andreas Dax[3], Henrik Lemke[3], Martin Decker[1], Nazaret Ortiz[1], Chris Milne[3], Grigory Smolentsev[8], Laura Maurel[5,6], Steven L. Johnson[7], Akihiro Mitsuda[9], Hirofumi Wada[9], Yuichi Yokoyama[10,11], Hiroki Wadati[10,11,12], and Urs Staub[1,‡]

[1] Swiss Light Source, Paul Scherrer Institute, CH-5232 Villigen PSI, Switzerland

[2] Deutsches Elektronen-Synchrotron DESY, Notkestraße 85, 22607, Hamburg, Germany

[3] SwissFEL, Paul Scherrer Institute, CH-5232 Villigen PSI, Switzerland

[4] Laboratoire de Spectroscopie Ultrarapide (LSU) and Lausanne Centre for Ultrafast Science (LACUS), Ecole Polytechnique Fédérale de Lausanne (EPFL), ISIC-FSB, Station 6, 1015 Lausanne, Switzerland

[5] Laboratory for Multiscale Materials Experiments, Paul Scherrer Institute, CH-5232 Villigen PSI, Switzerland

[6] Laboratory for Mesoscopic Systems, Department of Materials, ETH Zurich, 8093 Zurich, Switzerland

[7] Institute for Quantum Electronics, ETH Zürich, 8093 Zürich, Switzerland

[8] Paul Scherrer Institute, CH-5232 Villigen PSI, Switzerland

[9] Department of Physics, Kyushu University, Fukuoka 819-0395, Japan

[10] Institute for Solid State Physics, University of Tokyo, Chiba 277-8581, Japan

[11] Department of Physics, University of Tokyo, Tokyo 113-0033, Japan

[12] Graduate School of Material Science, University of Hyogo, 3-2-1, Koto, Kamigori-cho, Ako-gun, Hyogo 678-1297, Japan

## Abstract

**Ultrafast electron delocalization induced by a fs laser pulse is a well-known process and is the initial step for important applications such as fragmentation of molecules or laser ablation in solids. It is well understood that an intense fs laser pulse can remove several**

**electrons from an atom within its pulse duration. [1] However, the speed of electron localization out of an electron gas, the capture of an electron by ion, is unknown. Here, we demonstrate that electronic localization out of the conduction band can occur within only a few hundred femtoseconds. This ultrafast electron localization into 4f states has been directly quantified by transient x-ray absorption spectroscopy following photo-excitation of a Eu based correlated metal with a fs laser pulse. Our x-ray experiments show that the driving force for this process is either an ultrafast reduction of the energy of the 4f states, a change of their bandwidth or an increase of the hybridization between the 4f and the 3d states. The observed ultrafast electron localization process raises further basic questions for our understanding of electron correlations and their coupling to the lattice.**

Valence fluctuations in 4f electron systems have been studied intensively in the past, [2] particularly focusing on heavy fermion behavior, Kondo insulators, magnetic order, the occurrence of non BCS like superconductivity and non-Fermi liquid states. However, valence transitions and their fluctuations are still not well understood, even for a prototypical system such as elemental Cerium, which exhibits a first order valence transition as a function of pressure. [3] Many correlated metals are based on the fact that the f-electron shell is partly occupied with 4*f* states lying at the Fermi surface. Therefore, hybridization with the conduction electrons play a significant role in the electronic and magnetic properties.

Here, we focus on, the $EuNi_2(Si_{1-x}Ge_x)_2$ system, which exhibits a valence transition [4] that depends on temperature, [5] magnetic field, [6] on the substitutional level x [7] that acts as internal pressure [4], and on applied external pressure [8] [9]. While the Eu valence is closer to 3+ for temperatures below the phase transition, it is closer to 2+ above the phase transition. [5] The transition is further associated with a large change in the crystal lattice volume. Electrons localized at the Eu ion reduce the attractive potential of the ion, which increases its ionic radius, resulting in a very strong electron-lattice coupling. [5] Analogous to increasing temperature, the amount of the $Eu^{2+}$ ions (localized electrons), increases with increasing magnetic fields. [5] Despite $Eu^{3+}$ having a non-magnetic *J*=0 ground state, the Eu ions have a spin *S*=3 and angular momentum *L*=-3. Hence, magnetic excitation are observed throughout the valence transition in the $EuCu_2(Si/Ge)_2$ analogue. [10] The decrease of the Eu valence with increasing temperature is an interesting aspect for a valence transition as it is opposite to the effect observed in correlated metals such as, $YbAgCu_4$, [11] $YbInCu_4$ [12] or $SmB_6$,[13] where the Lanthanide valence increases with increasing temperatures. In the latter systems, these findings are interpreted as a strong reduction in hybridization between the *f* and *d* electrons at elevated temperatures.

Therefore, the EuNi$_2$(Si$_{1-x}$Ge$_x$)$_2$ system is an ideal case to test how fast electrons can localize from the conduction band. Exciting the material with an intense fs laser pulse must result in charge localization, at least on long timescales after which the electronic and lattice subsystems are equilibrated and the effective temperature has increased. A simple schematic of the electronic states is shown in Figure 1a. The valence instability is caused by localized 4*f* states that lie at the Fermi surface $E_F$. An increase of 4*f* level occupation (decrease of the Eu valence) can therefore be obtained by the following mechanisms: i) lowering the energy of the 4*f* states with respect to $E_F$ or ii) increasing the 4*f*-electron band width. Both processes can result in the transfer of conduction *s-d* electrons to localized *f* electrons. Recently, a light-induced valence transition has recently been confirmed by probing the valence of Eu ions 50ps after photo excitation with at the Eu M$_5$-edge soft x-ray absorption spectroscopy (XAS). [14] The estimated time scale of the valence change was found to be faster than 50 ps, which was limited by the time resolution of the experiment. As such, the origin and the time scale of the valence transition is still an open question.

We employ fs time resolved (tr-) Eu $L_3$-edge XAS, which is an extremely sensitive technique to quantify the timescale of this electron localization process and to provide a better understanding of the underlying mechanism. We chose a compound with x=0.79 for which the valence transition occurs at approximately 100K, which can be more conveniently accessed. The Eu$^{2+}$ and Eu$^{3+}$ $L_3$-edge resonances are separated by 7eV as demonstrated in Figure 1b. The tr-XAS experiments have been performed at the Bernina endstation [15] of SwissFEL in a pump probe setup as sketched in Figure 1c. The metallic sample was excited by 800 nm, 45fs (FWHM) laser pulses and the Eu valence state probed by x-rays with energies in the vicinity of the Eu$^{3+}$ (Eu$^{2+}$) resonances. An incident angle of 0.5 degrees was chosen to obtain an x-ray penetration depth of 30 nm, optimizing the overlap of the pumped and probed sample volume. The resulting x-ray fluorescence was detected by a Jungfrau detector. Simultaneously, the x-ray intensity of the (002) reflections was acquired with a second Jungfrau detector allowing us to determine the transient lattice dynamics. See the method section for further details.

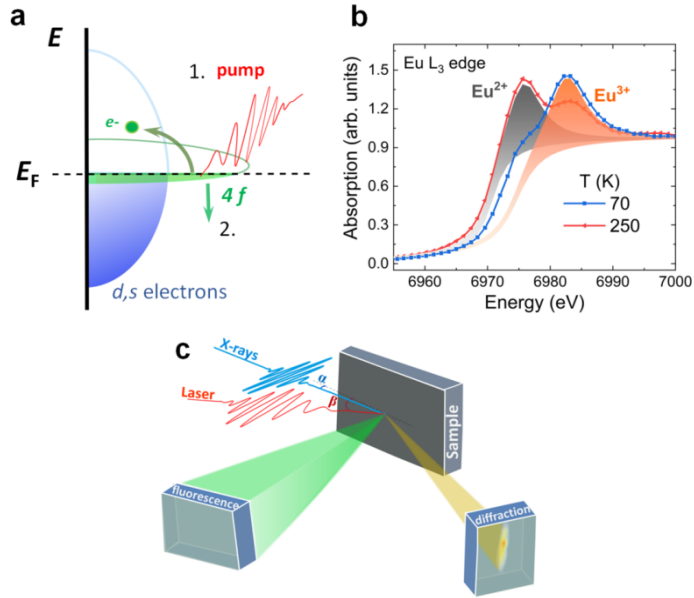

**Figure 1: a)** Schematic of the conduction band and modifications of the 4f states upon laser excitation. **b)** Eu $L_3$ edge XAS spectra below (blue) and above (red) the valence-transition temperature. The peaks located at 6.977 keV and 6.983 keV are assigned to the Eu $2p_{3/2} \rightarrow 5d$ dipole transition ($L_3$ edge) of $Eu^{2+}$ and $Eu^{3+}$ ions, respectively. **c)** Sketch of the time-resolved x-ray diffraction/absorption experimental setup. The incident angles α and β for the x-ray and laser are 0.5 and 5 degrees, respectively.

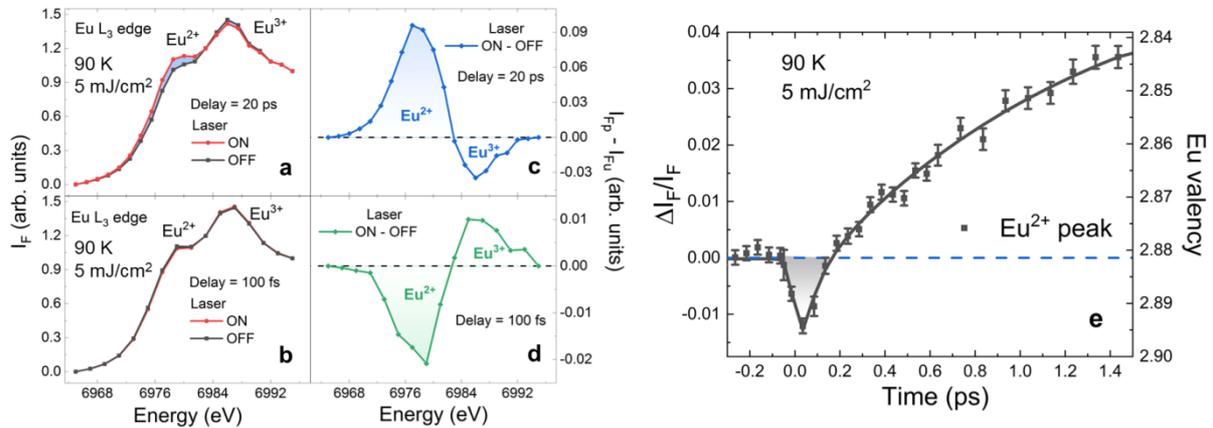

**Figure 2:** Time-resolved x-ray absorption data with energies around the Eu $L_3$ edge taken at 90 K. Panels a) and b) exhibit the normalized XANES spectra for $EuNi_2(Si_{0.21}Ge_{0.79})_2$ from the excited ($I_{Fp}$, laser on) and unperturbed ($I_{Fu}$, laser off) sample for time delays of a) 20 ps and b) 100 fs. The laser fluence was adjusted to 5 mJ/cm². Panels c) and d) show the difference between the unperturbed and transient absorption spectra at delays of 20 ps and 100 fs after excitation, respectively. Panel e) displays the normalized intensity [($I_{Fp}$-$I_{Fu}$)/$I_F$] collected at a fixed energy (at the $Eu^{2+}$ resonance of 6.977 keV) as a function of time delay.

The Eu $L_3$ tr-X-ray absorption near-edge structure (tr-XANES) spectra of $EuNi_2(Si_{0.21}Ge_{0.79})_2$ are shown in figures 2a and 2b for a 20 ps and 100 fs time delay between optical excitation and x-ray probe, respectively. A clear change of the $Eu^{2+}$ and $Eu^{3+}$ resonances becomes apparent when plotting the differences between the transient and the unperturbed spectra, with opposite sign for the two delay times (Figure 2c,d). A time delay scan performed at the energy corresponding to the $Eu^{2+}$ resonance is shown in Figure 2e. It clearly illustrates that the x-ray fluorescence reduces promptly after photo excitation, consistent with an initial enhancement of the Eu valence (getting closer to 3+). This indicates that during the excitation process, localized 4*f* electrons are transferred to the itinerant conduction band. Such an ultrafast electron delocalization is expected for a laser excitation.

However, shortly after this initial suppression, the fluorescence signal from the $Eu^{2+}$ ions begins to increase. The signal rises above its initial value after 200 fs, demonstrating an enhanced electron localization into 4*f* states. This directly represents an ultrafast electron localization process. Changes of the x-ray fluorescence at a given energy could also be due to ultrafast changes of the local atomic structure around the Eu ion represented by changes of the 5*d* final state crystal field splitting. Clearly, the changes in the XAS spectra at given time delays in figure 2c,d are fully consistent with the changes in the redox state observed upon simply increasing the temperature (see supplementary information), which excludes such a scenario.

The observed ultrafast charge localization raises the question about the nature of the intrinsic process causing it. In a first step, the various changes in the band structure consistent with the observed changes can be evaluated. (i) The 4*f* band energy could be lowered, (ii) the 4*f* bandwidth could be increased or (iii) the hybridization between the 4*f* band and the conduction electrons could be affected. The latter would result in a change of the character of electrons that become more or less *f*-like. To address these scenarios, we collected optical pump probe reflectivity data at 90K for various fluences with 800 nm pump and probe radiation (shown in Figure 3a). The optical reflectivity data show a sign change similar to the tr-XANES at a comparable fluence. The tr-XANES response (see Supplementary Figure S4) shows a slowing down of the charge localization process for increasing fluences. A similar behavior is also observed for the signal in the optical data (Figure 3a), with a higher signal to noise ratio. Using a simple phenomenological model (see SM), we extract the relaxation times of the positive signal from both the optical and the x-ray data. The resulting time constants are plotted as function of the excitation fluence in Figure 3b.

The relaxation time increases approximately linearly with increasing fluence, which is in strong contrast to what has been observed in intermediate valent systems with strong 4*f* - conduction band hybridization such as $YbInCu_4$ or $SmB_6$ [16], where the relaxation time strongly decreases for increasing fluence. This decrease for the latter systems has been interpreted in terms of a reduction of the 4*f* - conduction band hybridization. In other words, the relaxation time of these correlated metals approach the much faster response times of an uncorrelated metal when either the temperature or the pump fluence is increased. [16] EuNi$_2$(Si$_{0.21}$Ge$_{0.79}$)$_2$ contrasting behavior compared to YbInCu$_4$ or SmB$_6$ leads us to conclude that in our system the photo excitation increases the *f-d* (*s*) hybridization. This contrast in the photo-excitation response between these classes of materials is likely connected to the opposite temperature dependence of the Lanthanide valence between these materials. While the valence increases in YbInCu$_4$ or SmB$_6$ with increasing temperature, it decreases in the case of EuNi$_2$(Si$_{0.21}$Ge$_{0.79}$)$_2$.

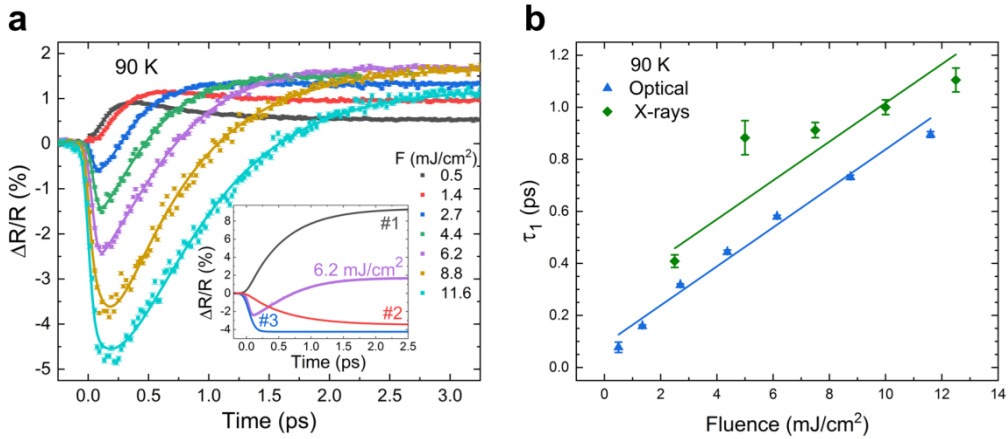

**Figure 3: Panel a) shows the transient 800 nm reflectivity as a function of fluence collected at 90 K. The inset shows the three exponential curves employed to fit the data. Panel b) shows the fluence dependence of the recovery time τ$_1$ obtained from the x-ray and optical measurements at 90 K. The solid green and blue lines represent linear fits.**

Having clarified that the changes of the electronic structure are caused by an increase of hybridization, it remains to explain why the electronic structure changes at all. Assuming a purely electronic effect (such as screening) to be at its origin, a much faster change of electron localization would be expected. [17] This is not consistent with the observed instantaneous process of electron delocalization during the laser excitation for the pump fluences used in the x-ray experiment. Note that the optical data might hint towards a threshold behavior, as fluences <= 1.4mJ/cm$^2$ do not show an instant drop in the reflectivity data. Independent of this signature, it is important to consider the induced lattice dynamics, in particular as the electron localization is strongly coupled to it. We expect three major processes that could be important: 1) lattice expansion, 2) coherent excitation of a specific phonon mode, or 3) coupling to random lattice fluctuations represented by Debye Waller terms measurable in x-ray diffraction. We therefore recorded simultaneously to the fluorescence XAS x-ray diffraction data of the (002) Bragg reflection. Figure 4a shows the (002) intensity projected to the two-theta axis for various excitation fluences. Clear changes in the peak position are observed (Figure 4 b), which can be directly transferred to changes of the unit cell volume (see Figure 4c) under the assumption that only the out of surface direction is affected by the excitation. This assumes that at early timescales the in-plane directions remain locked to the underlying undisturbed lattice. The data shows that the lattice dynamics show similar features as the electronic changes observed by XAS: An initial compression is followed by an expansion. However such a lattice dynamic is much slower and delayed compared to the change in valence. This decoupling proves that the electron localization drives the lattice motion and not vice versa.

The second scenario could also be tested by x-ray diffraction, as it would result in intensity changes of the Bragg peaks. No effect on the (002) reflection intensity is observed below 1 ps. Note that the expected changes on an intense low Q reflection are much smaller than the errors of the observed intensities preventing us from drawing any conclusions from the diffraction intensity data. However, the excitation of a coherent phonon mode with a frequency below 6 THz (frequency limited by the time resolution of the optical experiment) would most likely be visible in the optical data. While the optical data shows changes with similar timescale as observed in the XAS measurements, no coherent modulations are visible. In addition, considering the quartic response on the mode amplitude, as observed e.g. by ultrafast x-ray diffraction in cuprates [18] or pnictides [19], a fast initial change on the order of a quarter period of the mode frequency would be expected, which is inconsistent with our data.

The third scenario would lead to a clear decay of diffraction intensity due to the Debye-Waller term. Unfortunately, our data are inconclusive as this effect is expected to be very small for low Q reflections. However, the observed time scale is on the order of the time scale expected for the energy transfer between the electronic subsystem to the lattice, which suggests that lattice fluctuations are likely involved in the electron localization process.

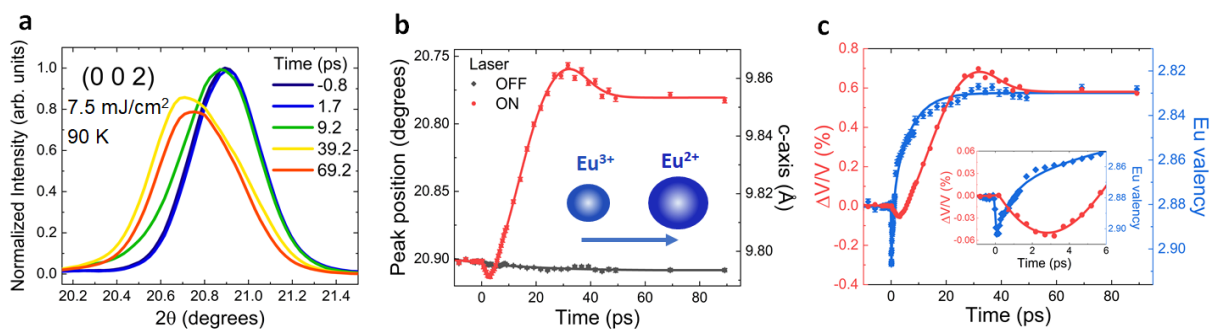

**Figure 4: a) (002) Bragg reflection intensities at various time delay after photo excitation with a fluence of 7.5 mJ/cm$^2$ taken at 90 K. b) Fitted (002) peak position and corresponding crystallographic c axis lattice constant as a function of time for perturbed (red symbols) and unperturbed system (grey symbols). The inset shows schematically the increase of the ionic radii between Eu$^{2+}$ ions and Eu$^{3+}$ after excitation. c) Relative expansion of the chemical unit cell along the c-axis and the Eu valency as a function of time. Inset: Zoom into short delay times of c).**

An incoherent interaction between the lattice and the electron localization brings up interesting considerations. First, there is a large angular magnetic moment change between the two Eu valence states. (See Supplementary Figure S5d) Since angular momentum is conserved, the electron localization process must transfer the momentum from/to the lattice. Ultrafast angular momentum transfer from the magnetization from a ferromagnetic Fe film to the lattice has recently been detected through ultrafast x-ray diffraction that observed the creation of an acoustic shear wave. [20] The transition from the Eu$^{3+}$ $J$=0, $L$=3 and $S$=-3 state to a $J$=7/2 $L$=0 state, requires the annihilation of a large angular magnetic moment at the Eu site that is randomly oriented due to the absence of magnetic order. A moment transfer to/from the lattice would result in an incoherent creation/annihilation of phonons. It is therefore possible that an angular magnetic moment transfer acts as a bottleneck for the electron localization. Another interesting point is the consideration that the initial state exhibits thermal fluctuations of the average nearest neighbor distance to the Eu. Hence one may expect that electrons are more likely localized at times the average distance around a selected ion is enhanced. The fact that a more intense photo excitation results in slower responses, but must result in larger excitation induced fluctuations, disagrees with this scenario. However, the initial state is in a valence fluctuating regime and possibly the fastest timescale at the lowest fluence is related to the time scale of the initial state valence fluctuations. Though our results show that the ultrafast electron localization of electrons out of the conduction band is correlated increase of hybridization and not driven by a coherent motion of the lattice, it also raises a series of fundamental questions for correlated electron systems and their dynamics. In particular the results point to the fact that the atomic angular momentum and the local structural distortion are important in this ultrafast electronic transition. Understanding their role will have impact on our general understanding of electron correlations and in physics of strongly correlated electron systems.

# Methods

**Sample preparation and characterization**

Polycrystalline EuNi$_2$(Si$_{0.21}$Ge$_{0.79}$)$_2$ samples have been grown by argon-arc melting and are described elsewhere. [21] An sample with dimensions of approximately 3 x 4 mm$^2$ (and m ~ 49 mg) was used. In order to obtain an optically shiny and flat surface and remove possible oxides (Eu$_2$O$_3$) from the surface, the sample was mechanically polished. Static x-ray absorption spectroscopy and magnetic susceptibility measurements to characterize the sample were performed at the SuperXAS-X10DA beam line at the Swiss Light Source and at the Laboratory for Multiscale Materials Experiments at PSI, respectively (Supplementary Information). Simultaneous time-resolved x-ray diffraction and XANES measurements were carried out at the Bernina instrument [15] of the SwissFEL facility using the optical pump/ X-ray probe technique. A Si(111) double crystal monochromator was used to tune the x-ray energy around the Eu L$_3$-edge (6977 eV). The x-ray beam was focused down to 30 μm x 110 μm by using a pair of Kirkpatrick-Baez (KB) mirrors and the free electron laser (FEL) repetition rate was either 10 or 25 Hz.

The experiment was performed around a temperature of 90K (below T$_S$ ~ 100 K). To reach this temperature, the sample was cooled with a Helium cryojet that was placed inside a chamber with controlled Helium atmosphere to avoid ice formation and to reduce x-ray scattering and absorption from the air. The temperature of the sample was monitored during the experiment with a sensor that has been calibrated using a Cernox temperature sensor positioned on the sample holder.

The sample was excited by 800 nm Ti-sapphire femtosecond laser with pulse duration of 45 fs FWHM, focal spot size of approximately 370 x 230 μm$^2$ and the fluence was varied between 2.5 and 12.5 mJ/cm$^2$ with the repetition rate half of that of the x-rays in order to alternately probe the pumped and unpumped sample. The angle of incidence of the 800 nm pump laser was 5°. The incoming x-ray beam interacts with the sample at grazing angle of 0.5° in order to reduce the x-ray penetration depth to 30 nm.

Three Jungfrau detectors [22] were used to collect 2D images for each FEL shot. X-ray scattering images were recorded with a 1.5M pixel detector. Another 0.5M pixel detector collected the XANES signal and was placed at 90 degrees two theta angle to minimize elastically scattered x-rays. A third 0.5M pixel Jungfrau served as an I0 intensity monitor by collecting the scattering from a Capton foil placed into the incoming x-ray beam.

For each fluence and for selected photon energies, at least two time scans were collected from -5 to 30 ps with 1000 shots per time delay. The data were filtered for very low and high intensity shots and then normalized by I0. See supplementary materials for more information on the data treatment.

* Authors with equal contributions

‡ [urs.staub@psi.ch](mailto:urs.staub@psi.ch)

## DATA AVAILABILITY

The datasets generated during and/or analysed during the current study, including analysing scripts are available from the corresponding authors on reasonable request.

## ACKNOWLEDGMENTS

We like to acknowledge Kojiro Mimura Valerio Scagnoli and Maarten Nachtegaal for informative discussions. This work was supported by the NCCR Molecular Ultrafast Science and Technology (NCCR MUST), a research instrument of the Swiss National Science Foundation (SNSF). Time-resolved X-ray measurements were carried out on the Bernina endstation at the SwissFEL facility, Paul Scherrer Institut (PSI), Switzerland. X-ray absorption characterization experiments were performed on the SuperXAS beamline at the Swiss Light Source (SLS), PSI. Optical pump-probe data were measured on the FEMTO-microXAS laser system at SLS, PSI.

This work was supported by MEXT Quantum Leap Flagship Program (MEXT Q-LEAP) Grant No. JPMXS0118068681 and by JSPS KAKENHI Grant No. 19H05824. We also acknowledge support of the European Research Council Advanced Grants H2020 ERCEA 695197 DYNAMOX

….

## AUTHOR CONTRIBUTIONS

U.S. and J.R.L.M. conceived the experiment; J.R.L.M., S.Z., G.M., V.E., J.R., R.M., C.S., N.G., S.P., M.P., B.B., D.Y., P.B., G.I., B.P., C.A., C.E., A.D., H.L., S.L.J., and U.S. carried out the X-ray pump-probe experiments; J.R.L.M., S.Z., V.E., J.R. performed the optical measurements; S.Z. and V.E. wrote the online data analysis code; S.Z. performed the pump-probe data analysis; J.R.L.M. and S.Z. performed the final data analysis; G.M. designed the experiment at SwissFEL; J.R.L.M., N.G., B.B., N.O., G.S., L.M. performed the static sample characterization; A.M., H.W., Y.Y. and H.W. provided the sample and performed further characterization measurements; U.S. wrote the manuscript with input from all authors.

**ADDITIONAL INFORMATION**

Supplementary information available online (url to be inserted during production)

**Competing interests**: The authors declare no competing Financial or Non-Financial interests.